\documentclass[draftclsnofoot,12pt, onecolumn]{IEEEtran}
\usepackage{amsmath}
\usepackage{epsfig}
\usepackage{amssymb}
\usepackage{latexsym}
\usepackage{graphicx}
\usepackage{cite}

\title{On the $\eta-\mu$/gamma and the $\lambda-\mu$/gamma \\ Composite Distributions}
\author{\IEEEauthorblockN{Paschalis C. Sofotasios\\}
\IEEEauthorblockA{School of Electronic and Electrical Engineering  \\
University of Leeds, UK\\
Email: p.sofotasios@leeds.ac.uk\\}
\and
\IEEEauthorblockN{Steven Freear\\}
\IEEEauthorblockA{School of Electronic and Electrical Engineering\\
University of Leeds, UK\\
Email: s.freear@leeds.ac.uk}}
\begin{document}
\maketitle
\begin{abstract} 
This work is devoted to the formulation and derivation of the $\eta{-}\mu{/}$gamma and $\lambda{-}\mu{/}$gamma distributions which correspond to physical fading models. These distributions are composite and are based on the $\eta-\mu$ and $\lambda-\mu$ generalized multipath models, respectively, and the gamma shadowing model. Novel analytic expressions are derived for the corresponding envelope probability density functions. Importantly, the proposed models provide accurate characterisation of the simultaneous occurrence of multipath fading and shadowing effects which is achieved thanks to the remarkable flexibility offered by their parameters that render them capable of providing good fittings to experimental data associated with realistic communication scenarios. This is additionally justified by the fact that they include as special cases the widely known fading models such as Hoyt/gamma, Nakagami-m/gamma and Rayleigh/gamma. As a result, they can be meaningfully utilized in various analytical studies related to the performance evaluation of digital communications over composite multipath/shadowing fading channels.
\end{abstract}

%
%
%
\section{Introduction}
\indent
It is widely accepted that fading is a physical phenomenon that degrades communication signals during wireless propagation. A common approach for accounting for this effect is to exploit appropriate statistical distributions which are typically  known as \textit{fading models}. As a result, statistical models such as Rayleigh, Nakagami-$m$, Weibull and Nakagami-q (Hoyt) have been shown to be capable of modelling small-scale fading in Non-Line-of-Sight (NLOS) communication scenarios, whereas the Nakagami-$n$ (Rice) distribution has been typically utilized in characterizing multipath fading in Line-of-Sight (LOS) communication scenarios, \cite{B:Nakagami, B:Jakes, B:Alouini} and the references therein. Capitalizing on the these models, M. D. Yacoub initially proposed three generalised fading distributions, namely, the $\alpha-\mu$, the $\kappa-\mu$, the $\eta-\mu$ models and subsequently the $\lambda-\mu$ and the $\kappa-\mu$ Extreme models, \cite{C:Yacoub_1, C:Yacoub_2, C:Yacoub_3, J:Yacoub_1, J:Yacoub_2, C:Yacoub_4, C:Rabelo}. These models are distinct thanks to the remarkable flexibility offered by their named parameters which render them capable of providing adequate fitting to experimental data which correspond to realistic communication scenarios. Their usefulness is also evident by the fact that they include as special cases all the aforementioned small-scale fading distributions. \\
\indent
However, it is recalled that a crucial aspect of wireless radio propagation is that the occurrence of multipath and shadowing effects is typically simultaneous. As a consequence, in spite of the undoubted usefulness of the aforementioned fading models, they all ultimately fail to account concurrently for both shadowing and multipath fading. In other words, the utilization of the aforementioned fading models is limited to the effective characterisation of either the one or the other effect. Based on this principal limitation, the need for composite statistical models which are capable of providing efficient characterization of fading "as a whole", became evident \cite{B:Nakagami, B:Jakes, B:Alouini}.\\
\indent
To this effect, the authors in \cite{J:Kaveh} proposed the Rayleigh/gamma fading model, which is also known as $K$-distibution, $(K)$. Likewise, Shankar in \cite{J:Shankar} exploited the flexibility of Nakagami-$m$ distribution, which includes Rayleigh distribution as a special case, and introduced the Nakagami-$m$/gamma composite distribution - or generalised $K$-distribution, $(K_{G})$.  In the same context, the Weibull/gamma composite distribution was proposed in \cite{J:Bithas} while an introduction to more generalized composite distributions  was reported in \cite{C:Sofotasios_1, C:Sofotasios_2, C:Sofotasios_3, B:Sofotasios, Add_1, Add_2}. \\
\indent
Capitalizing on the above, the aim of this work is the formulation and derivation of the $\eta$-$\mu$/gamma, and the $\lambda$-$\mu$/gamma composite fading distributions. Specifically, after formulating these models, novel analytic expressions are derived for their corresponding envelope probability density function (pdf).  The validity of the offered results is justified numerically and their behaviour is examined under different parametric scenarios. Thanks to their relatively convenient algebraic representation, the offered results are considered useful mathematical tools that can be efficiently utilized in analytical studies related to the performance of digital communications over $\eta-\mu$/gamma and $\lambda-\mu$/gamma composite fading channels. To this effect, they can be meaningfully exploited in the derivation of analytic expressions for critical performance metrics such as error probability, channel capacity and higher order statistics, among others.\\
\indent
The remainder of this paper is organised as follows: Section II revisits the basic principles of the $\eta-\mu$, the $\lambda-\mu$ and the gamma distributions. Sections III and IV are devoted to the presentation, formulation and derivation of the $\eta-\mu$/gamma and $\lambda-\mu$/gamma fading models, respectively, along with the necessary analysis on their behaviour. Finally, discussions on their potential applicability in wireless communications along with closing remarks are given in Section V.  
%
%
\section{The $\eta-\mu$, $\lambda-\mu$ and gamma Fading Distributions}
\subsection{The $\eta-\mu$ Fading Model}
\indent
The $\eta-\mu$ distribution is a general fading distribution that accounts for the small-scale variation of fading signals in NLOS communication scenarios. It is written in terms of two physical parameters, $\eta$ and $\mu$; the former corresponds to the ratio of the powers between the multipath waves in the in-phase and quadrature components, whereas the latter is related to the number of multipath clusters in the environment \cite{C:Yacoub_1, J:Yacoub_2}.
For a fading signal with envelope $R$ and $\hat{r} = \sqrt{E(R^{2})}$, the $\eta-\mu$ envelope probability density function (pdf) is mathematically expressed as

\begin{equation} \label{eq:1} 
p_{_{R}}(r) = \frac{4 \sqrt{\pi}\mu^{\mu + \frac{1}{2}} h^{\mu}}{\hat{r}\,\Gamma(\mu)H^{\mu - \frac{1}{2}}} \left(\frac{r}{\hat{r}} \right)^{2\mu} e^{-2\mu h \left(\frac{r}{\hat{r}} \right)^{2}} I_{\mu - \frac{1}{2}}\left[ 2\mu H \left( \frac{r}{\hat{r}}\right)^{2} \right]
\end{equation}
where $E(.)$ and $\hat{r}$ denote expectation and the root-mean-square $(rms)$ value of $R$, respectively. Furthermore, the parameter $\mu$ is positive and is given by 

\begin{equation} \label{eq:2} 
\mu = \frac{E^{2}(R^{2})}{2Var(R^{2})}\left[1 + \left(\frac{H}{h} \right) \right]
\end{equation}
where $var(.)$ denotes variance and the parameters $h$ and $H$ are defined as, 

\begin{equation} \label{eq:3} 
h = \frac{2 + \eta^{-1} +\eta}{4}
\end{equation}
and

\begin{equation} \label{eq:4} 
H = \frac{\eta^{-1} - \eta}{4}
\end{equation}
Also $0< \eta< \infty$ is as already mentioned the scattered wave power ratio between the in-phase and quadrature components of each cluster of multipath. By taking the ratio of $h$ and $H$, it is shown that \cite{J:Yacoub_2}

\begin{equation} \label{eq:5} 
 \frac{H}{h} = \frac{1 - \eta}{1 + \eta}
\end{equation}
Equation \eqref{eq:1} can be also expressed in a normalized form by considering a normalized envelope $P=R/\hat{r}$. Based on this, the corresponding normalized power pdf can be deduced by setting $p_{W}(w) = p_{P}(\sqrt{w})/2\sqrt{w}$, namely,

\begin{equation} \label{eq:6} 
p_{W}(w) = \frac{2 \sqrt{\pi}\mu^{\mu + \frac{1}{2}} h^{\mu}}{\Gamma(\mu)H^{\mu - \frac{1}{2}}}w^{\mu - \frac{1}{2}} e^{-2\mu h w} I_{\mu - \frac{1}{2}}\left( 2\mu Hw\right)
\end{equation}
\indent
It is recalled that the $\eta-\mu$ model includes as special cases other widely known small-scale fading distributions. More specifically, the Hoyt distribution is obtained for $\mu = 0.5$ and $b = (1 - \eta)/(1+\eta)$ (where $b$ denotes the Nakagami-$q$ or Hoyt parameter). Likewise, the one sided Gaussian distribution is attained for $\eta \rightarrow \infty$ or $\eta \rightarrow 0$ whereas the Nakagami-$m$ distribution is attained for $\mu = m$ and $\eta \rightarrow 0$ or $\eta \rightarrow \infty$. Finally, the $\eta-\mu$ distribution reduces to the Rayleigh distribution for  the special case $\mu=0.5$ and $\eta = 1$, \cite{J:Yacoub_2}.
\subsection{The $\lambda-\mu$ Fading Model}
\indent
The $\lambda-\mu$ distribution is a fading model which constitutes the \textit{Format}-$2$ of the $\eta-\mu$ distribution in \cite{J:Yacoub_2}. Recalling the same initial conditions as in the previous Subsection, its normalized envelope pdf is obtained by setting in \eqref{eq:1}, 

\begin{equation} \label{eq:7} 
h = \frac{1}{1 - \eta^{2}}
\end{equation}
and

\begin{equation} \label{eq:8} 
H = \frac{\eta}{1 - \eta^{2}}
\end{equation}
Therefore, for $\eta = \lambda$, it follows that

\begin{equation} \label{eq:9} 
\lambda = \frac{H}{h}
\end{equation}
which is valid for \, $-1<\lambda <1$. To this effect, the corresponding pdf is given by \cite[eq. (1)]{C:Yacoub_4}, namely, 

\begin{equation} \label{eq:10} 
p_{_{P}}(\rho) = \frac{4\sqrt{\pi}\mu^{\mu + \frac{1}{2}} \rho^{2\mu}}{\Gamma(\mu)\lambda^{\mu - \frac{1}{2}}\sqrt{1 - \lambda^{2}}} e^{-\frac{2\mu \rho^{2}}{1 - \lambda^{2}}}I_{\mu - \frac{1}{2}}\left(\frac{2\mu\lambda}{1 - \lambda^{2}} \rho^{2}\right) 
\end{equation}
where

\begin{equation} \label{eq:11} 
\mu = \frac{E^{2}(R^{2})}{2Var(R^{2})}\left(1 + \lambda \right)
\end{equation}
By utilizing once more the relationship $p_{W}(w) = p_{P}(\sqrt{w})/2\sqrt{w}$, its normalized power pdf is expressed as

\begin{equation} \label{eq:12} 
p_{_{W}}(w) = \frac{2\sqrt{\pi}\mu^{\mu + \frac{1}{2}} w^{\mu - \frac{1}{2}}}{\Gamma(\mu)\lambda^{\mu - \frac{1}{2}}\sqrt{1 - \lambda^{2}}} e^{-\frac{2\mu w}{1 - \lambda^{2}}}I_{\mu - \frac{1}{2}}\left(\frac{2\mu\lambda}{1 - \lambda^{2}} w\right) 
\end{equation}
\indent
Physically, likewise the $\eta-\mu$ model (Format $1$), the parameter $\mu$ is  the inverse of the normalised variance and relates to the number of multipath clusters. On the contrary, the parameter $\lambda$, or $\eta$ in Format $2$ of the $\eta-\mu$ model, denotes the correlation coefficient between the scattered wave in-phase and quadrature components of each cluster of multipath \cite{J:Yacoub_2}. Importantly, the Nakagami-$m$ fading distribution is obtained for $\mu = m$ and $\lambda \pm 1$, whereas for the special case $\mu=0.5$ and $\lambda = 1$, the $\lambda-\mu$ distribution reduces to the Rayleigh distribution \cite{C:Yacoub_4}.

\subsection{The Gamma Fading Model}
\indent
The log-normal distribution has been largely considered the optimum statistical model for characterising the shadowing effect, \cite{B:Nakagami, B:Jakes, B:Alouini}. Nevertheless, in spite of its usefulness, it has been largely shown that when it becomes algebraically involved with other elementary and/or special function, its algebraic representation often renders it inconvenient to handle analytically. This is particularly the case in studies related to the analytical derivation of critical performance measures in digital communications over fading channels. Motivated by this, the authors in \cite{J:Kaveh} proposed the gamma distribution as an accurate substitute to log-normal distribution. Mathematically, the envelope pdf of gamma distribution is given by \cite[eq. (4)]{J:Kaveh}, namely,

\begin{equation} \label{13} 
p_{_{Y}}(y) = \frac{y^{b-1}e^{-\frac{y}{\Omega}}}{\Gamma(b) \Omega^{b}}, \qquad \, \, y\geq 0 
\end{equation}
where the term $b > 0$ is its shaping parameter and $\Omega = E(Y^{2})$. This fading model has been shown to provide adequate fitting to experimental data that correspond to realistic fading conditions. In addition, its algebraic representation is particularly tractable and therefore, easy to handle both analytically and numerically. As a result, it has been widely considered a useful model for characterising shadowing effect and based on this, it has been exploited in the formulation of the $K$ and $K_{G}$ composite multipath/shadowing models, \cite{J:Kaveh, J:Shankar}.
%
%
\section{The $\eta-\mu$/gamma Fading Distribution}

\subsection{Model Formulation}
\indent
According to the basic principles of statistics, the envelope pdf of a composite statistical distribution is constituted by superimposing two or more statistical distributions. In the present case, this is realized by superimposing one multipath and one shadowing distribution, namely,

\begin{equation} \label{14} 
p_{_{R}}(r) \triangleq \int_{0}^{\infty} p_{_{R\mid Y}}(r\mid y)p_{_{Y}}(y)dy
\end{equation}
where $p_{_{R\mid Y}}(x\mid y)$ denotes the corresponding multipath distribution with mode $y$. Evidently, the $\eta-\mu$/gamma composite fading distribution is formulated by firstly setting $r = x$ and $\hat{r}=y$ in \eqref{eq:1} and then substituting  in \eqref{14} along with equation \eqref{13}. To this end, it immediately follows that

\begin{equation} \label{15} 
p_{_{X}}(x) = \frac{4\sqrt{\pi}\mu^{\mu + \frac{1}{2}}h^{\mu} x^{2\mu}}{\Gamma(\mu) H^{\mu - \frac{1}{2}}\Gamma(b) \Omega^{b}}\int_{0}^{\infty} \frac{e^{-\frac{y}{\Omega}} \,I_{\mu - \frac{1}{2}}\left(2\mu H \frac{x^{2}}{y^{2}} \right)}{y^{ 2(\mu + 1) - b}\,e^{2\mu h \frac{x^{2}}{y^{2}}}}  dy
\end{equation}
Importantly, the term $y^{2}$ in \eqref{15} has emerged from the term $\hat{r}^{2}$ which denotes that the mean-squared value of the fading amplitude follows the gamma distribution. However,  it is noted here that it can be also assumed that the root-mean-squared value of the fading amplitude is gamma distributed. In fact, this is exactly the difference between the Rayleigh/Lognormal and Suzuki fading models since in the former the $rms$ value of the fading amplitude is modelled as by log-normal distribution, whereas in the latter it is the mean-squared value of this amplitude which is assumed to be log-normally distributed \cite{B:Alouini}. Therefore, by applying this principle in \eqref{15} and letting $u=1/y$, it follows that $y = 1/u$ and $du/dy = -1/y^{2}$. To this effect and by taking the corresponding absolute value, the following expression is deduced, 

\begin{equation} \label{16} 
p_{_{X}}(x) = \frac{4\sqrt{\pi}\mu^{\mu + \frac{1}{2}}h^{\mu} x^{2\mu}}{\Gamma(\mu) H^{\mu - \frac{1}{2}}\Gamma(b) \Omega^{b}}\int_{0}^{\infty} \frac{e^{-\frac{1}{\Omega u}} \,I_{\mu - \frac{1}{2}}\left(2\mu H x^{2}u \right)}{u^{b - \mu +\frac{1}{2}}\,e^{2\mu h x^{2}u}}  du
\end{equation}

\subsection{The Special Case $\mu \in \mathbb{N} $ }

A closed-form expression for the envelope pdf of the $\eta-\mu$/gamma distribution for the case that $\mu \in \mathbb{N}$ can be obtained with the aid of the closed-form series representation for the $I_{n}(x)$ function in \cite[eq. (8.467)]{B:Tables}, namely,

\begin{equation} \label{17} 
I_{n + \frac{1}{2}} (x)\triangleq \sum_{k=0}^{n}\frac{(n+k)!\,\left[(-1)^{k}e^{x} + (-1)^{n+1}e^{-x}\right]}{\sqrt{\pi}k!(n-k)!(2x)^{k+\frac{1}{2}}}, \, n\in \mathbb{N}
\end{equation}
As a result, by performing the necessary change of variables, recalling that $\Gamma(x)\triangleq (x-1)!$ and substituting in \eqref{16} yields,

$$
p_{_{X}}(x) = \sum_{k=0}^{\mu -1} \frac{\Gamma(\mu +k)\mu^{\mu - k}h^{\mu}x^{2(\mu - k) - 1}}{k!\Gamma(\mu - k)2^{2k + 1}\Gamma(\mu)\Gamma(b)H^{\mu + k}\Omega^{b}}  \qquad \qquad \qquad \qquad \qquad \qquad \qquad \qquad \qquad \qquad 
$$
\begin{equation} \label{18} 
\times \left\lbrace (-1)^{k}\int_{0}^{\infty}u^{\mu - b - k - 1}e^{-\frac{1}{\Omega u}}e^{-2\mu x^{2} (h-H)u}du + (-1)^{\mu}\int_{0}^{\infty}u^{\mu - b - k - 1}e^{-\frac{1}{\Omega u}}e^{-2\mu x^{2} (h+H)u}du\right\rbrace 
\end{equation}
 
The above integrals can be solved in closed-form with the aid of \cite[eq. (3.471.9)]{B:Tables}, namely,

\begin{equation} \label{19} 
\int_{0}^{\infty}x^{\nu - 1}e^{-\frac{b}{x}}e^{-\gamma x}dx = 2\left(\frac{b}{\gamma} \right)^{\frac{\nu}{2}}K_{\nu}\left(2\sqrt{b\gamma} \right)
\end{equation}
where $K_{\nu}(x)$ denotes the modified Bessel function of the second kind, \cite{B:Abramowitz}. Therefore, by performing the necessary variable transformation and substituting in \eqref{18}, one obtains

$$
p_{_{X}}(x) = S\delta(x) + \sum_{k=0}^{\mu - 1} \frac{2^{2 - \frac{\mu + 3k -b}{2}}\mu^{\frac{\mu +b -k}{2}}(\mu)_{k}h^{\mu}x^{\mu -k +b -1}}{k!\Gamma(m-k)\Gamma(b)H^{\mu +k}\Omega^{b + \frac{\mu -k-b}{2}}} \times 
$$   
\begin{equation} \label{20} 
\left\lbrace \frac{K_{ \mu -b-k}\left(2x\sqrt{\frac{2\mu(h-H)}{\Omega}} \right)}{(-1)^{-k}(h-H)^{\frac{ \mu -b-k}{2}}} + \frac{K_{ \mu -b-k }\left(2x\sqrt{\frac{2\mu(h+H)}{\Omega}} \right)}{(-1)^{-\mu}(h+H)^{\frac{ \mu -b-k}{2}}} \right\rbrace 
\end{equation}
where $\delta(.)$ is the Dirac delta function and $(x)_{n} \triangleq \Gamma(x+n)/\Gamma(x)$ is the Pochhammer symbol \cite{B:Tables}. Furthermore, the parameter $S$ denotes a normalisation scalar constant that must be determined so that equation \eqref{20} constitutes a true pdf. To this end, by recalling that $\int_{0}^{\infty}p_{_{X}}(x) dx \triangleq 1$ and $\int_{0}^{\infty}\delta(x)dx \triangleq 1$, one obtains,

$$
S = 1 - \sum_{k=0}^{\mu -1} \frac{2^{2 - \frac{\mu + 3k -b}{2}}\mu^{\frac{\mu +b -k}{2}}(\mu)_{k}h^{\mu}}{k!\Gamma(m-k)\Gamma(b)H^{\mu +k}\Omega^{b + \frac{\mu -k-b}{2}}} \qquad  \qquad \qquad \qquad \qquad \qquad \qquad \qquad \qquad \qquad 
$$
\begin{equation} \label{21} 
\left\lbrace \frac{(-1)^{k}}{(h-H)^{\frac{\mu -b -k}{2}}}\int_{0}^{\infty}\frac{K_{\mu - b- k}\left(2x\sqrt{\frac{2\mu(h-H)}{\Omega}} \right)}{x^{1 + k - \mu -b}}dx  + \frac{(-1)^{\mu}}{(h+H)^{\frac{\mu -b -k}{2}}} \int_{0}^{\infty}\frac{K_{\mu -b -k }\left(2x\sqrt{\frac{2\mu(h+H)}{\Omega}} \right)}{x^{1 + k -\mu -b}} dx\right\rbrace
\end{equation}
Importantly, the above integrals can be solved with the aid of \cite[eq. (6.561.16)]{B:Tables}, namely, 

\begin{equation} \label{22} 
\int_{0}^{\infty}x^{a}K_{n}(bx)dx = \frac{2^{a-1}}{b^{a+1}}\Gamma\left(\frac{a +1 +n}{2} \right)\Gamma\left(\frac{a +1 -n}{2} \right)
\end{equation}
which is valid for $ \mathrm{Re}\, b>0$ and $\mathrm{Re}\, a +1 \pm n>0$. To this effect, by making the necessary change of variables and substituting in \eqref{21} yields the following relationship,

\begin{equation} \label{23} 
S = 1 - \sum_{k=0}^{\mu -1}\frac{(\mu)_{k} h^{\mu}}{k!(2H)^{\mu +k}} \left[\frac{(-1)^{k}}{(h-H)^{\mu - b -k}} + \frac{(-1)^{\mu}}{(h+H)^{\mu -b -k}} \right]
\end{equation}
Therefore, by substituting \eqref{23} into \eqref{21}, a closed-form expression for the envelope pdf of the $\eta$-$\mu$/gamma fading model for $\mu \in \mathbb{N}$ is finally deduced, namely,

$$
p_{_{X}}(x) = \sum_{k=0}^{\mu - 1} \frac{2^{2 - \frac{\mu + 3k -b}{2}}\mu^{\frac{\mu +b -k}{2}}(\mu)_{k}h^{\mu}x^{\mu -k +b -1}}{k!\Gamma(m-k)\Gamma(b)H^{\mu +k}\Omega^{b + \frac{\mu -k-b}{2}}} \left\lbrace \frac{K_{ \mu -b-k}\left(2x\sqrt{\frac{2\mu(h-H)}{\Omega}} \right)}{(-1)^{-k}(h-H)^{\frac{ \mu -b-k}{2}}} + \frac{K_{ \mu -b-k }\left(2x\sqrt{\frac{2\mu(h+H)}{\Omega}} \right)}{(-1)^{-\mu}(h+H)^{\frac{ \mu -b-k}{2}}} \right\rbrace 
$$   
 
\begin{equation} \label{24} 
- \sum_{k=0}^{\mu -1}\frac{(\mu)_{k} h^{\mu}}{k!(2H)^{\mu +k}} \left[\frac{(-1)^{k}}{(h-H)^{\mu - b -k}} + \frac{(-1)^{\mu}}{(h+H)^{\mu -b -k}} \right] + 1
\end{equation}
\subsection{The general case $\mu \in \mathbb{R}$}
\indent
By recalling that the the root-mean-squared value - and not the mean-squared value- of the fading amplitude is gamma distributed, equation \eqref{15} is alternatively expressed as, 

\begin{equation} \label{25} 
p_{_{X}}(x) = \frac{4\sqrt{\pi}\mu^{\mu + \frac{1}{2}}h^{\mu} x^{2\mu}}{\Gamma(\mu) H^{\mu - \frac{1}{2}}\Gamma(b) \Omega^{b}}\int_{0}^{\infty} \frac{e^{-\frac{y}{\Omega}} \,I_{\mu - \frac{1}{2}}\left(2\mu H \frac{x^{2}}{y} \right)}{y^{\frac{3}{2} + \mu - b}\,e^{2\mu h \frac{x^{2}}{y}}}  dy
\end{equation}
Notably, the modified Bessel function can be also expressed in terms of the polynomial approximation in \cite[eq. (19)]{J:Gross} as follows:

\begin{equation} \label{26} 
I_{\nu}(x) \simeq \sum_{l = 0}^{n} \frac{\Gamma(n + l)}{\Gamma(l + 1) \Gamma(n - l + 1)}\frac{n^{1 - 2l}}{\Gamma(\nu + l + 1)} \left(\frac{x}{2} \right)^{\nu + 2l}.
\end{equation}
As $n \rightarrow \infty$, the above expression reduces to the infinite series in \cite[eq. (8.445)]{B:Tables}. Therefore, by making the necessary variable transformation and substituting in \eqref{26}, one obtains

\begin{equation} \label{27} 
p_{_{X}}(x) = \sum_{k=0}^{n}\frac{ \Gamma(n + k)\mu^{2(\mu + k)}H^{2k}h^{\mu}n^{1-2k}x^{4(\mu +k)-1}\Omega^{-b}}{\Gamma(k+1)\Gamma(n-k+1)\Gamma\left(\mu + k + \frac{1}{2} \right)\Gamma(\mu)\Gamma(b)} 4\sqrt{\pi} \int_{0}^{\infty} y^{b - 2\mu -2k -1}e^{-\frac{y}{\Omega}}e^{-\frac{2\mu h x^{2}}{y}}dy
\end{equation}
By utilizing again \eqref{19} in \eqref{27} and after long algebraic manipulations, the following closed-form expression is deduced,

\begin{equation}\label{28} 
p_{_{X}}(x) = S\delta(x) + \sum_{k=0}^{n} \frac{2^{\frac{b}{2}-\mu - k +3}\sqrt{\pi}\Gamma(n+k)\mu^{\frac{b}{2} + \mu +k}H^{2k}}{k!\Gamma(n-k+1)\Gamma\left(\mu +k + \frac{1}{2}\right)\Gamma(\mu)}  \left\lbrace \frac{n^{1-2k} x^{2(\mu + k)+b -1}}{h^{k - \frac{b}{2}}\Gamma(b)\Omega^{\frac{b}{2}+\mu +k}}K_{b - 2(\mu +k)}\left(2x\sqrt{\frac{2\mu h}{\Omega}}\right)\right\rbrace
\end{equation}
The scalar parameter $S$ that constitutes $p_{_{X}}(x)$ a true pdf can be determined by following the same methodology as in the previous Subsection. To this end, it is noticed that the corresponding resulted integral has the same algebraic form as the integral in \eqref{22}. Based on this, by making the necessary change of variables, substituting in \eqref{28} and recalling that $\int_{0}^{\infty} p(x)dx = \int_{0}^{\infty}\delta(x)dx = 1$ yields the following expression

\begin{equation}\label{29} 
S = 1 - \sum_{k=0}^{n} \frac{\sqrt{\pi}\Gamma(n+k)H^{2k}n^{1-2k}\Gamma(2\mu +2k)2^{1 -2(\mu +k) }}{k!\Gamma(n-k+1)\Gamma\left(\mu + k +\frac{1}{2}\right)\Gamma(\mu)h^{\mu +2k}}.
\end{equation}
As a result, by substituting the above expression into \eqref{28}, one obtains the following analytic expression for $\mu \in \mathbb{R}$, 

\begin{equation}\label{30} 
\begin{split}
p_{_{X}}(x) =& 1 - \sum_{k=0}^{n} \frac{\sqrt{\pi}\Gamma(n+k)H^{2k}n^{1-2k}\Gamma(2\mu +2k)2^{-2(\mu +k) }}{2k!\Gamma(n-k+1)\Gamma\left(\mu + k +\frac{1}{2}\right)\Gamma(\mu)h^{\mu +2k}}\\
&+ \sum_{k=0}^{n} \frac{2^{\frac{b}{2}-\mu - k +3}\sqrt{\pi}\Gamma(n+k)\mu^{\frac{b}{2} + \mu +k}H^{2k}}{k!\Gamma(n-k+1)\Gamma\left(\mu +k + \frac{1}{2}\right)\Gamma(\mu)} \left\lbrace \frac{n^{1-2k} x^{2(\mu + k)+b -1}}{h^{k - \frac{b}{2}}\Gamma(b)\Omega^{\frac{b}{2}+\mu +k}}K_{b - 2(\mu +k)}\left(2x\sqrt{\frac{2\mu h}{\Omega}}\right)\right\rbrace 
\end{split}
\end{equation}
To the best of the authors' knowledge, expressions in \eqref{24} and \eqref{30}, have not been previously reported in the open literature. 
%
%
\section{The $\lambda-\mu$/gamma Fading Distribution}
\subsection{Model Formulation}
\indent
As already mentioned in Section I, the $\lambda-\mu$ fading model is a different version of the $\eta-\mu$ fading model (format 2). As a result, the algebraic form of the corresponding envelope pdf of the two distributions is the same and therefore, analytic expressions for the $\lambda-\mu$/gamma composite model can be derived based in the same manner as the $\eta-\mu$/gamma model in Section III. To this end, by assuming that the $rms$ of the fading amplitude follows the gamma distribution, setting $r=x$ in \eqref{eq:10} and substituting it in \eqref{14} along with \eqref{eq:11}, yields an explicit expression for the envelope pdf of the $\lambda-\mu$/gamma distribution, namely

\begin{equation} \label{31} 
p_{_{X}}(x) = \frac{4\sqrt{\pi}\mu^{\mu + \frac{1}{2}}\lambda^{\frac{1}{2} - \mu}x^{2\mu}}{\Gamma(\mu) \sqrt{1 - \lambda^{2}}\Gamma(b) \Omega^{b}}\int_{0}^{\infty} \frac{e^{-\frac{y}{\Omega}} \,I_{\mu - \frac{1}{2}}\left(\frac{2\mu \lambda}{1-\lambda^{2}} \frac{x^{2}}{y} \right)}{y^{\frac{3}{2} + \mu - b}\,e^{\frac{2\mu}{1 - \lambda^{2}}  \frac{x^{2}}{y}}}  dy
\end{equation}

\subsection{The special case $\mu \in \mathbb{N}$}
\indent
 By representing the $I_{n}(x)$ function in \eqref{31} according to \eqref{17} and setting $u = 1/y$, one obtains the following expression,

$$
p_{_{X}}(x) = \sum_{k=0}^{\mu -1} \frac{\Gamma(\mu +k)\mu^{\mu - k}\left(1 - \lambda^{2} \right)^{k} x^{2(\mu - k) - 1}}{k!\Gamma(\mu - k)2^{2k + 1}\Gamma(\mu)\Gamma(b)\lambda^{\mu +k}\Omega^{b}}  \qquad \qquad \qquad   \qquad \qquad \qquad   \qquad \qquad \qquad   \qquad \qquad \qquad  
$$
\begin{equation} \label{32} 
\times \left\lbrace (-1)^{k}\int_{0}^{\infty}u^{\mu - b - k - 1}e^{-\frac{1}{\Omega u}}e^{-\frac{2\mu x^{2}(1 - \lambda)}{1 - \lambda^{2}} u}du + (-1)^{\mu}\int_{0}^{\infty}u^{\mu - b - k - 1}e^{-\frac{1}{\Omega u}}e^{-\frac{2\mu x^{2}(1 + \lambda)}{1 - \lambda^{2}} u}du\right\rbrace 
\end{equation}
which has the same algebraic representation as \eqref{18}. Therefore, by performing the necessary change of variables and making use of \eqref{19}, the following expression for $p_{_{X}}(x)$  is deduced,

$$
p_{_{X}}(x) = S\delta(x) + \sum_{k=0}^{\mu - 1} \frac{4\mu^{\frac{\mu +b -k}{2}}(\mu)_{k} \left(1 - \lambda^{2} \right)^{\frac{\mu +k -b}{2}} x^{\mu -k +b -1}}{2^{\frac{\mu + 3k -b}{2}} k!\Gamma(m-k)\Gamma(b)\lambda^{\mu +k}\Omega^{\frac{\mu -k-b}{2}}} 
$$   
\begin{equation} \label{33} 
\times \left\lbrace \frac{K_{ \mu -b-k}\left(2x\sqrt{\frac{2\mu(1-\lambda)}{\Omega \left(1 - \lambda^{2}\right)}} \right)}{\Omega^{b}(-1)^{-k}(1 - \lambda)^{\frac{ \mu -b-k}{2}}} + \frac{K_{ \mu -b-k }\left(2x\sqrt{\frac{2\mu(1 + \lambda)}{\Omega \left(1 - \lambda^{2}\right)}} \right)}{\Omega^{b}(-1)^{-\mu}(1 +\lambda)^{\frac{ \mu -b-k}{2}}} \right\rbrace 
\end{equation}
Next, by following the same procedure as in Section III, the scalar parameter $S$ is given by,

 $$
S = 1 - \sum_{k=0}^{\mu -1} \frac{2^{2 - \frac{\mu + 3k -b}{2}}\mu^{\frac{\mu +b -k}{2}}(\mu)_{k}\left(1 - \lambda^{2} \right)^{\frac{\mu +k -b}{2}}}{k!\Gamma(m-k)\Gamma(b)\lambda^{\mu +k}\Omega^{b + \frac{\mu -k-b}{2}}} \times \qquad  \qquad  \qquad  \qquad  \qquad  \qquad  \qquad  \qquad  \qquad  \qquad  
$$

\begin{equation} \label{34} 
\left\lbrace \frac{(-1)^{k}}{(1 - \lambda)^{\frac{\mu -b -k}{2}}}\int_{0}^{\infty}\frac{K_{\mu - b- k}\left(2x\sqrt{\frac{2\mu(1 - \lambda)}{\Omega \left(1 - \lambda^{2} \right)}} \right)}{x^{1 + k - \mu -b}}dx  + \frac{(-1)^{\mu}}{(1 + \lambda)^{\frac{\mu -b -k}{2}}} \int_{0}^{\infty}\frac{K_{\mu -b -k }\left(2x\sqrt{\frac{2\mu(1 +\lambda)}{\Omega \left(1 - \lambda^{2} \right)}} \right)}{x^{1 + k -\mu -b}} dx\right\rbrace
\end{equation}
Importantly, the above two integrals belong to the same class as the integral in \eqref{22}. Therefore, by performing the necessary variable transformation it follows straightforwardly that,

\begin{equation} \label{35} 
S = 1 - \sum_{k=0}^{\mu -1}\frac{(\mu)_{k} \left(1 - \lambda^{2} \right)^{k+1}}{k!(2\lambda)^{\mu +k}} \left[\frac{(-1)^{k}}{(1 -\lambda)^{\mathcal{A}  -k}} + \frac{(-1)^{\mu}}{(1 +\lambda)^{\mathcal{A}-k}} \right]
\end{equation}
where $\mathcal{A} = \mu -b$. Evidently, by substituting \eqref{35} into \eqref{34} yields a closed-form expression for the envelope pdf of the $\lambda-\mu$ fading model for $\mu \in \mathbb{N}$, namely,

$$
p_{_{X}}(x) = \sum_{k=0}^{\mu - 1} \frac{4\mu^{\frac{\mu +b -k}{2}}(\mu)_{k} \left(1 - \lambda^{2} \right)^{\frac{\mu +k -b}{2}} x^{\mu -k +b -1}}{2^{\frac{\mu + 3k -b}{2}} k!\Gamma(\mu-k)\Gamma(b)\lambda^{\mu +k}\Omega^{\frac{\mu -k-b}{2}}} \left\lbrace \frac{K_{ \mu -b-k}\left(2x\sqrt{\frac{2\mu(1-\lambda)}{\Omega \left(1 - \lambda^{2}\right)}} \right)}{\Omega^{b}(-1)^{-k}(1 - \lambda)^{\frac{ \mu -b-k}{2}}} + \frac{K_{ \mu -b-k }\left(2x\sqrt{\frac{2\mu(1 + \lambda)}{\Omega \left(1 - \lambda^{2}\right)}} \right)}{\Omega^{b}(-1)^{-\mu}(1 +\lambda)^{\frac{ \mu -b-k}{2}}} \right\rbrace 
$$

\begin{equation} \label{36} 
+1 - \sum_{k=0}^{\mu -1}\frac{(\mu)_{k} \left(1 - \lambda^{2} \right)^{k+1}}{k!(2\lambda)^{\mu +k}} \left[\frac{(-1)^{k}}{(1 -\lambda)^{\mathcal{A}  -k}} + \frac{(-1)^{\mu}}{(1 +\lambda)^{\mathcal{A}-k}} \right] 
\end{equation}
\begin{figure}[h]
\centerline{\psfig{figure=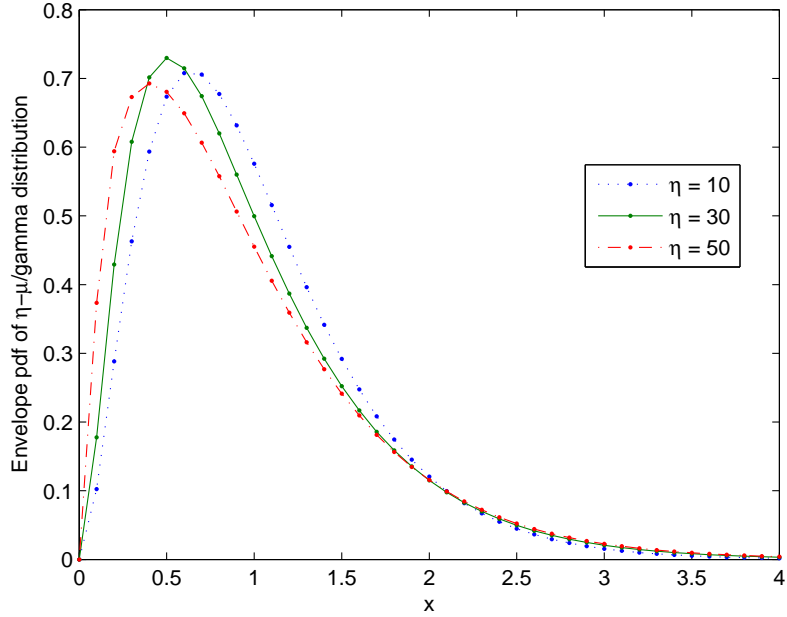, width=12cm, height=9cm}}
\caption{Envelope pdf of the $\eta$-$\mu$/gamma distribution for $b=1.2$, $\Omega=0.8$, $\mu = 0.6$ and different values of $\eta$.}
\end{figure}
\subsection{The general case $\mu \in \mathbb{R}$}
\indent
An explicit expression for the envelope pdf of the $\lambda-\mu$/gamma distribution when $\mu \in \mathbb{R}$, can be derived with the aid of the polynomial approximation for the $I_{n}(x)$ in \eqref{26}. To this effect, by making the necessary variable transformation and inserting in \eqref{31} yields

$$
p_{_{X}}(x) = \sum_{k=0}^{n}\frac{ \Gamma(n + k)\mu^{2(\mu + k)} \lambda^{2k} n^{1-2k}x^{4(\mu +k)-1}}{k!\Gamma(n-k+1) \left(1 - \lambda^{2} \right)^{\mu + 2k} \Gamma\left(\mu + k + \frac{1}{2} \right) }
$$
\begin{equation} \label{37} 
\times \frac{4\sqrt{\pi}}{\Gamma(\mu)\Gamma(b)\Omega^{b}} \int_{0}^{\infty} y^{b - 2\mu -2k -1}e^{-\frac{y}{\Omega}}e^{-\frac{2\mu x^{2}}{y \left(1 - \lambda^{2} \right)}}dy
\end{equation}
Evidently, the above relationship can be expressed in explicit form by evaluating the involved integral. Likewise in previous Sections, the algebraic representation of this integral is the same as the integral in \eqref{19}. As a result, the following analytic expression is deduced in a straightforward manner,

$$
p_{_{X}}(x) = S\delta(x) + \sum_{k=0}^{n} \frac{2^{\frac{b}{2}-\mu - k +3}\sqrt{\pi}\Gamma(n+k)\mu^{\frac{b}{2} + \mu +k}\lambda^{2k}}{k!\Gamma(n-k+1)\Gamma\left(\mu +k + \frac{1}{2}\right)\Gamma(\mu)} \times
$$
\begin{equation}\label{38} 
\left\lbrace \frac{n^{1-2k} x^{2(\mu + k)+b -1}}{\left(1 - \lambda^{2} \right)^{k + \frac{b}{2}}\Gamma(b)\Omega^{\frac{b}{2}+\mu +k}}K_{b - 2(\mu +k)}\left(2x\sqrt{\frac{2\mu }{\Omega \left(1 - \lambda^{2} \right)}}\right)\right\rbrace
\end{equation}
The scalar normalization constant  $S$ needs to be also determined in order to constitute $p_{_{X}}(x)$ a true pdf, i.e. $\int_{0}^{\infty}p_{_{X}}(x)dx = 1$. To this end, by integrating both parts of \eqref{38} from zero to infinity, recalling that $\int_{0}^{\infty}\delta(x)dx = 1$ and utilizing \eqref{22} yields the following closed-form expression

\begin{equation}\label{39} 
S = 1 - \sum_{k=0}^{n} \frac{2\sqrt{\pi}\Gamma(n+k)n^{1-2k} \lambda^{2k}\left(1 - \lambda^{2} \right)^{\mu} \Gamma(2\mu +2k)}{2^{2(\mu +k) } k!\Gamma(n-k+1)\Gamma\left(\mu + k +\frac{1}{2}\right)\Gamma(\mu)}.
\end{equation}
Finally, by substituting \eqref{39} in \eqref{38} yields an analytic relationship for the envelope pdf of the $\lambda-\mu$/gamma fading distribution for the case of $\mu \in \mathbb{R}$, namely, 

$$
p_{_{X}}(x) = 1 + \sum_{k=0}^{n} \frac{2^{\frac{b}{2}-\mu - k +3}\sqrt{\pi}\Gamma(n+k)\mu^{\frac{b}{2} + \mu +k}\lambda^{2k}}{k! \Gamma(b)\Gamma(n-k+1)\Gamma\left(\mu +k + \frac{1}{2}\right)\Gamma(\mu)} \times
$$
$$
\left\lbrace \frac{n^{1-2k} x^{2(\mu + k)+b -1}}{\left(1 - \lambda^{2} \right)^{k + \frac{b}{2}}\Omega^{\frac{b}{2}+\mu +k}}K_{b - 2(\mu +k)}\left(2x\sqrt{\frac{2\mu }{\Omega \left(1 - \lambda^{2} \right)}}\right)\right\rbrace
$$
\begin{equation} \label{40} 
 - \sum_{k=0}^{n} \frac{\sqrt{\pi}\Gamma(n+k)n^{1-2k} \lambda^{2k}\left(1 - \lambda^{2} \right)^{\mu} \Gamma(2\mu +2k)}{2^{2(\mu +k) -1} k!\Gamma(n-k+1)\Gamma\left(\mu + k +\frac{1}{2}\right)\Gamma(\mu)}
\end{equation}
To the best of the authors' knowledge, the proposed expressions in equations \eqref{36} and \eqref{40} are novel. 
%
\begin{figure}[h]
\centerline{\psfig{figure=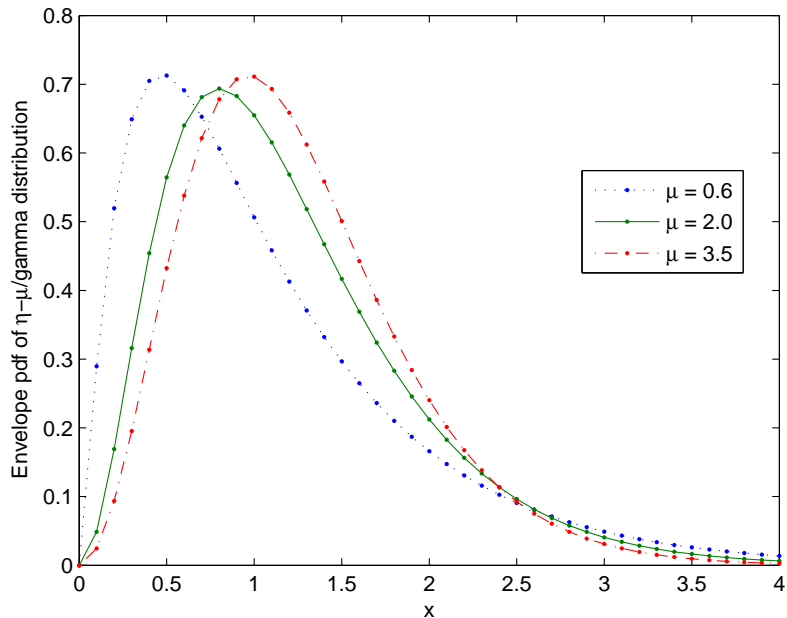,width=12cm, height=9cm}}
\caption{Envelope pdf of the $\eta-\mu$/gamma distribution for $b=1.2$, $\Omega=0.8$, $\eta = 10$ and different values of $\mu$}
\end{figure}
\section{Numerical Results and Discussions}
\indent
In this Section, the general behaviour of the derived analytic expressions for the envelope pdf of the $\eta-\mu$/gamma and $\lambda-\mu$/gamma fading distributions is demonstrated. To this end, Figure $1$ illustrates the pdf of $\eta-\mu$/gamma with respect to $x$ for $b=1.2$, $\Omega=0.8$, $\mu = 0.6$ and different values of $\eta$. Likewise, Figure $2$ considers $b=1.2$, $\Omega=0.8$, $\eta = 0.6$ and different values of $\mu$. In the same context, Figures $3$ and $4$ demonstrate the pdf of $\lambda-\mu$/gamma for $b=\Omega=1.0$, $\mu = 0.6$ and different values of $\lambda$ and $b=1.25$, $\Omega=1.5$, $\lambda = 0.5$ and different values of $\mu$, respectively. One can observe the flexibility of the proposed models which render them capable of providing adequate fittings to experimental results. 
\begin{figure}[h]
\centerline{\psfig{figure=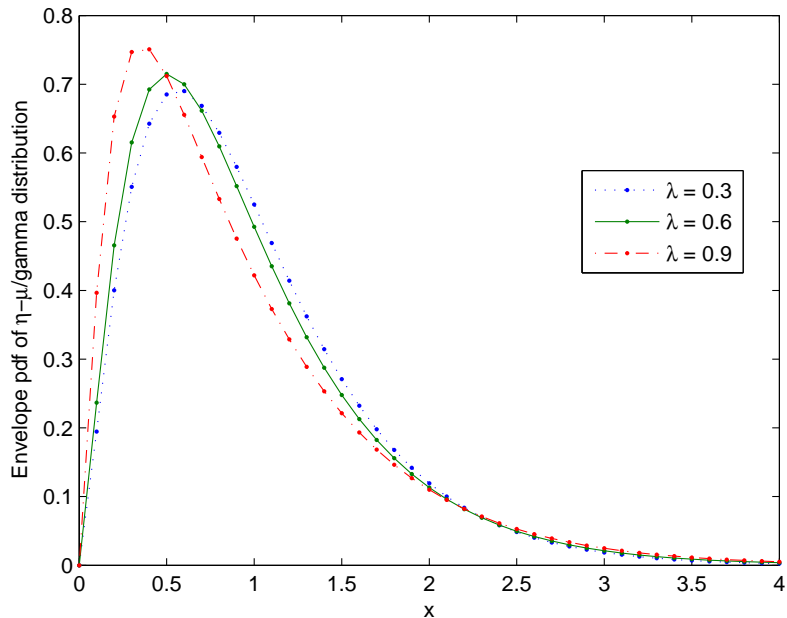,width=12cm, height=9cm}}
\caption{Envelope pdf of the $\eta-\mu$/gamma distribution for $b=\Omega=1.0$, $\mu = 0.6$ and different values of $\lambda$}
\end{figure}
\subsection{Usefulness and Applicability in Wireless Communications}
It is widely known that the algebraic representation of crucial performance measures is critical in studies related to analytical performance evaluation of digital communications. This is obvious by the fact that when the algebraic form of a corresponding measure is convenient, significantly increases the possibility that the derived relationships be expressed in closed-form. Therefore, the fact that the form of the offered analytic result have a relatively simple representation, renders this model convenient to handle both analytically and numerically. To this effect, the derived expressions can be efficiently applied in various analytic studies relating to the performance evaluation of digital communications over composite multipath/shadowing fading channels. Indicatively, the offered expressions can be straightforwardly utilized in deriving explicit expressions for important performance measures such as, error probability, probability of outage, ergodic capacity, channel capacity under different constraints and higher order statistics. It is recalled here that expressions corresponding to the aforementioned measures can obviously be derived in both classical and emerging technologies such as SISO and MIMO systems, diversity systems, cognitive radio and cooperative systems and optical communications among others. 
\begin{figure}[h]
\centerline{\psfig{figure=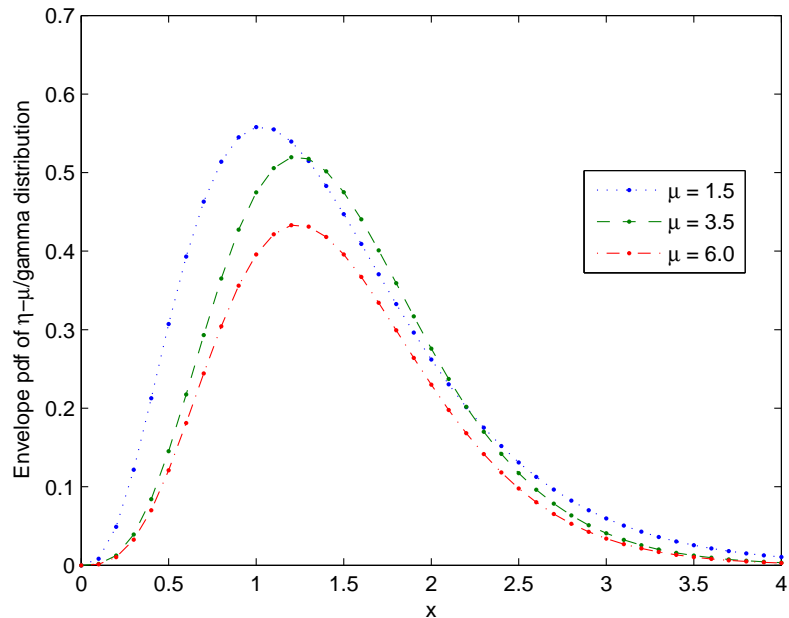,width=12cm, height=9cm}}
\caption{Envelope pdf of the $\eta-\mu$/gamma distribution for $b=\Omega=1.0$, $\lambda = 0.5$ and different values of $\mu$}
\end{figure}
\section{Closing Remarks}
\indent
This work was devoted in the introduction, formulation and derivation of the $\eta-\mu$/gamma and $\lambda-\mu$/gamma fading distributions. These models are formulated from the superimpose of the $\eta-\mu$ and $\lambda-\mu$ generalised small-scale fading models, respectively, and the gamma shadowing model. These distribution are particularly flexible and they include as special cases the  widely known Hoyt, Nakagami-$m$ and Rayleigh fading models. Novel analytic expressions were derived for the envelope probability density function which can be considered a useful mathematical tool in applications related analytical performance evaluation of digital communications over multipath/shadowing channels. 
\bibliographystyle{IEEEtran}

\end{document}